\documentstyle[12pt]{article}
\begin{document}
\begin{flushright}
{\large\bf Submit to $<<$Mod. Phys. Lett. B$>>$}
\end{flushright}
\parskip=6pt	
\baselineskip=20pt
\smallskip
\centerline{\Large\bf New Approach on the General Shape Equation} 
\centerline{\Large\bf of Axisymmetric Vesicles}
\bigskip
\bigskip
\centerline{\bf Zhan-Ning Hu\footnote{\bf E-mail: 
huzn@aphy.iphy.ac.cn}}
\centerline{Institute of Physics and Center 
for Condensed Matter Physics,} 
\centerline{Chinese Academy of Sciences, Beijing 100080, China}

\vspace{2ex}
\vspace{2ex}

\bigskip
\begin{center}
\begin{minipage}{5in}
\centerline{\large\bf 	Abstract} 
\vspace{1ex}
The general Helfrich shape equation determined by minimizing the 
curvature free energy describes the equilibrium shapes of the 
axisymmetric lipid bilayer vesicles in different conditions. 
It is a non-linear differential equation with  variable 
coefficients. In this letter, by analyzing the unique 
property of the solution, we change  this shape equation 
into a system of the two  differential equations. One of 
them is a linear differential equation. This equation system 
contains all of the known  rigorous solutions of the general 
shape equation.  And the more general constraint conditions 
are found for the solution of the general shape equation. 

\bigskip

\end{minipage}
\end{center}
\newpage

In recent years much attention has been payed on the investigations of 
the equilibrium shape of lipid bilayer vesicles in aqueous solution by 
physicists theoretically and experimentally [1-10].  The amphiphilic 
molecules such as phospholipids assemble to form bilayers in water and 
at low concentration close to take single shells which are called 
vesicles \cite{Szo}. These bilayer vesicles are considered as the 
models 
for biomembranes or cells \cite{wlE}. The  elastic property of the  
phospholipid vesicles is related to their behaviors about their 
formation, stability, size, fusion and budding processes. The 
Helfrich theory \cite{1111} and general shape equation \cite{6666} 
play an important role in the study of the equilibrium shapes of 
lipid bilayer vesicles. The equilibrium shape is determined by 
minimizing the curvature free energy \cite{1111}
\begin{equation}
F=\frac{k_c}{2}\oint(c_1+c_2-c_0)^2dA+\lambda\int dA+\Delta p\int dV
\end{equation}
where $k_c, c_1, c_2$ and $ c_0$ are  bending rigidity, two principal 
curvatures and  spontaneous curvature, respectively. $\lambda$ and 
$\Delta p$ are Lagrange multipliers due to the constraints of 
constant volume and area and the former denote a tensile stress 
and the later is the osmotic pressure difference between  outer 
and inner media. The general equation of mechanical equilibrium 
for  vesicle membranes is a non-linear differential equation 
with  variable coefficients, which is derived from the above 
curvature free energy. As the view   in Ref. \cite{naito}, 
the general shape equation of the axisymmetric vesicles is 
a nonlinear third-order differential equation and has the form:
$$ 
\cos^3\psi\Bigg(\frac{d^3\psi}{d\rho^3}\Bigg)=4\sin\psi  
\cos^2\psi\Bigg(\frac{d^2\psi}{d\rho^2}\Bigg)
\Bigg(\frac{d\psi}{d\rho}\Bigg)
-\cos\psi\Bigg(\sin^2\psi-\frac{1}{2}\cos^2\psi\Bigg)
\Bigg(\frac{d\psi}{d\rho}\Bigg)^3
$$
$$
+\frac{7\sin\psi  \cos^2\psi}{2\rho}
\Bigg(\frac{d\psi}{d\rho}\Bigg)^2
-\frac{2\cos^3\psi}{\rho}\Bigg(\frac{d^2\psi}{d\rho^2}\Bigg)
$$$$
+\Bigg[\frac{c^2_0}{2}-
\frac{2c_0\sin\psi}{\rho}+\frac{\lambda}{k}
-\frac{\sin^2\psi-2\cos^2\psi}{2\rho^2}\Bigg]\cos\psi
\Bigg(\frac{d\psi}{d\rho}\Bigg)
$$
\begin{equation}
+\Bigg[\frac{\Delta p}{k}+\frac{\lambda \sin\psi}{k\rho}+
\frac{c^2_0\sin\psi}{2\rho}
-\frac{\sin^3\psi+2\sin\psi \cos^2\psi}{2\rho^3}\Bigg],
\end{equation}
where $\psi$ is the angle between the surface normal and the 
symmetric axis, and $\rho$ is the distance from 
the symmetric axis. 

The very interesting work by Naito, Okuda and Ou-Yang \cite{naito} 
presented a new exact solution of general shape 
equation of axisymmetric vesicles, which enriches 
the solution family of  the shape equation. 
Now we have known that this general shape 
equation has the  solutions of spheres \cite{6666}, circular 
cylinders, Clifford torus \cite{8888}, Delaunay's surfaces, 
nodoidlike shapes  and  unduloidlike shapes \cite{naito}. These 
solutions were obtained by guessing and their correctness were 
checked by taking them into the shape equation. Are there any 
method to solve this nonlinear third order differential 
equation  for finding its solutions? It is just one of 
the aims in this letter. And we find that all the above 
shapes of vesicles are the particular solutions of a linear 
second order differential equation. The more general 
constraint conditions exist for the ``three-term'' solution 
comparing with the one in Ref. \cite{naito}. In the following, 
we discuss them in details.

Setting
\begin{equation}
\Psi=\sin\psi(\rho),
\end{equation}
we can write down the general shape equation as
$$
\big(\Psi^2-1\big)\frac{d^3\Psi}{d\rho^3}+\Psi\frac{d^2\Psi}
{d\rho^2}\frac{d\Psi}{d\rho}-\frac{1}{2}\Bigg(\frac{d\Psi}
{d\rho}\Bigg)^3+\frac{2(\Psi^2-1)}{\rho}\frac{d^2\Psi}
{d\rho^2}+\frac{3\Psi}{2\rho}\Bigg(\frac{d\Psi}{d\rho}
\Bigg)^2~~~~~~
$$
\begin{equation}
~~+\Bigg(\frac{c_0^2}{2}-\frac{2c_0\Psi}{\rho}+
\frac{\lambda}{k}-\frac{3\Psi^2-2}{2\rho^2}\Bigg)
\frac{d\Psi}{d\rho}+\frac{\Delta p}{k}+\Bigg(
\frac{\lambda}{k}+\frac{c_0^2}{2}-\frac{1}{\rho^2}
\Bigg)\frac{\Psi}{\rho}+\frac{\Psi^3}{2\rho^3}=0.
\end{equation}
The power series solutions for vesicles were studied 
in the interesting work \cite{ZOY} where they expanded 
the solution of the shape equation as the $infinite$ sum 
of a power series for the distance from the rotational   
$z$-axis.  In order to discuss the property of the cutoff 
of the power series solution we now put the solution 
of the general shape equation into the form 
\begin{equation}
\Psi=\sum^{M}_{s=m}a_s\rho^s.
\end{equation}
By an analysis of the shape equation we decide the 
parameters $m$ and $M$ in the above expression 
as follows. 
The terms related coefficient $a_s^3 (|s|>1)$ are
$$
\big(\Psi^2-1\big)\frac{d^3\Psi}{d\rho^3}\sim(s-2)(s-1)s
a^3_s\rho^{3(s-1)},~~\Psi\frac{d^2\Psi}{d\rho^2}
\frac{d\Psi}{d\rho}\sim (s-1)s^2a^3_s\rho^{3(s-1)},~~~~
$$$$
-\frac{1}{2}\Bigg(\frac{d\Psi}{d\rho}\Bigg)^3\sim -
\frac{1}{2}s^3a_s^3\rho^{3(s-1)}, ~~ 
\frac{2(\Psi^2-1)}{\rho}\frac{d^2\Psi}{d\rho^2}
\sim 2(s-1)sa^3_s\rho^{3(s-1)},~~~~~~~~
$$$$
\frac{3\Psi}{2\rho}\Bigg(\frac{d\Psi}{d\rho}
\Bigg)^2\sim \frac{3}{2}s^2a_s^3\rho^{3(s-1)}, ~~ 
\Bigg(\frac{c_0^2}{2}-\frac{2c_0\Psi}{\rho}+
\frac{\lambda}{k}-\frac{3\Psi^2-2}{2\rho^2}\Bigg)
\frac{d\Psi}{d\rho}\sim -\frac{3}{2}sa_s^3\rho^{3(s-1)},
$$
\begin{equation}
\frac{\Delta p}{k}+\Bigg(\frac{\lambda}{k}+\frac{c_0^2}{2}-
\frac{1}{\rho^2}\Bigg)\frac{\Psi}{\rho}+
\frac{\Psi^3}{2\rho^3}\sim\frac{1}{2}a_s^3\rho^{3(s-1)},~~~~
~~~~~~~~~~~~~~~~~
\end{equation} 
in the shape equation (4).
It gives that
\begin{equation}
(s-1)(s+1)(3s-1)=0.
\end{equation}
This means that $M=1, m=-1$ and expression (5) is
\begin{equation}
\Psi =\frac{a_{-1}}{\rho}+a_0+a_1\rho.
\end{equation}
We call it as the ``three-term'' solution 
of the general shape equation.  
On the side, from the Ref. \cite{7777}, we know that 
the shape equation (4) has the first integral form. 
So it can be written as 
\begin{equation}  \label{RRR}
(R^2\rho^2-1)\rho\frac{d^2R}{d\rho^2}-\frac{1}{2}R\rho^3
\Big(\frac{dR}{d\rho}\Big)^2+(2R^2\rho^2-3)
\frac{dR}{d\rho}-c_0R^2\rho+(\frac{\lambda}{k}+
\frac{c^2_0}{2}\Big)R\rho=f(\rho),
\end{equation}
where $R=\Psi/\rho$ and the function $f(\rho)$ 
satisfies the differential equation
\begin{equation}   
\frac{d}{d\rho}f(\rho)+\frac{1}{\rho}f(\rho)
+\frac{\Delta p}{k}=0,
\end{equation}
which gives that 
\begin{equation}   \label{fff}
f(\rho)=-\frac{\Delta p}{2k}\rho+\frac{C}{\rho}
\end{equation}
where $C$ is a integral constant. In this way, the general 
shape equation becomes as a nonlinear second-order 
differential equation:
\begin{equation} \label{YYY}
(Y^2-\rho^2)\frac{d^2Y}{d\rho^2}-\frac{Y}{2}
\Big(\frac{dY}{d\rho}\Big)^2+\rho\frac{dY}{d\rho}-c_0Y^2+
\Big(\frac{\lambda}{k}+\frac{c_0^2}{2}\Big)Y\rho^2+
\frac{\Delta p}{2k}\rho^4-C\rho^2=0
\end{equation}
with the use of the notation $Y=\rho^2R=\rho \Psi$. By 
considering the discussions in (6-8), that is, 
$M=1, m=-1$, and the known solutions given in 
Refs. \cite{6666,7777,naito}, 
we make the Ansatz:
\begin{equation}
Y^2\frac{d^2Y}{d\rho^2}-\frac{Y}{2}
\Big(\frac{dY}{d\rho}\Big)^2-c_0Y^2+
\Big(\frac{\lambda}{k}+\frac{c_0^2}{2}\Big)Y\rho^2+
\frac{\Delta p}{2k}\rho^4-C\rho^2-C_1-C_2\rho-C_3\rho^2=0,
\end{equation}
where $C_1, C_2$ and $C_3$ are the arbitrary constants. 
Then, from equation (12), we have that
\begin{equation}
\rho^2\frac{d^2Y}{d\rho^2}-\rho\frac{dY}{d\rho}
=C_1+C_2\rho+C_3\rho^2.
\end{equation}
It is a linear differential equation of the second 
order and can be changed into the first order one:
\begin{equation}
\frac{\widetilde{Y}}{d\rho}-\frac{1}{\rho}
\widetilde{Y}=\frac{C_1}{\rho^2}+\frac{C_2}
{\rho}+C_3,
\end{equation}
where $\widetilde{Y}=\displaystyle\frac{dY}
{d\rho}$. Then we get solution
\begin{equation}
Y=\frac{1}{2}(C_3\rho^2-C_1)\ln\rho+
\frac{1}{2}(A_1-\frac{C_3}{2})\rho^2-C_2\rho+A_2
\end{equation}
where $A_1, A_2$ are the integration 
constants. By substituting  
$$
\frac{dY}{d\rho}=C_3\rho\ln\rho+A_1\rho-C_2-\frac{C_1}{2\rho},
$$$$
\frac{d^2Y}{d\rho^2}=C_3\ln\rho+C_3+A_1+\frac{C_1}{2\rho^2},
$$
and relation (16)  into equation (13), we can 
get the constraint conditions of the physical 
parameters $\lambda, \Delta p, c_0$ and the 
constants $C_1, C_2, C_3, A_1, A_2$:
$$
2\rho^4(\ln\rho)^2C_3^2(2c_0-C_3)+\rho^4\ln\rho C_3
\Big(8A_1c_0-4A_1C_3-4c_0^2-4c_0C_3+3C_3^2-
\frac{8\lambda}{k}\Big)~~~
$$$$
+\rho^4\Big(4A_1^2c_0-2A_1^2C_3-4A_1c_0^2-4A_1c_0C_3+3A_1C_3^2-
\frac{8A_1\lambda}{k}+2c_0^2C_3+c_0C_3^2-C_3^3+
\frac{4C_3\lambda}{k}
$$$$
-\frac{8\Delta p}{k}\Big)+4\rho^3\ln\rho C_2C_3(-
4c_0+3C_3)+4\rho^3C_2
\Big(-4A_1c_0+3A_1C_3+2c_0^2+2c_0C_3-2C_3^2+
\frac{4\lambda}{k}\Big)
$$$$
-8\rho^2(\ln\rho)^2A_2C_3^2+4\rho^2\ln\rho C_3(-4A_1A_2+
4A_2c_0-2A_2C_3+C_2^2)+2\rho^2\Big(-4A_1^2A_2+8A_1A_2c_0
$$$$
-4A_1A_2C_3+2A_1C_2^2-4A_2c_0^2-4A_2c_0C_3+4A_2C_3^2-
\frac{8A_2\lambda}{k}+8C+8c_0C_2^2-9C_2^2C_3+8C_3\Big)
$$$$
+16\rho\ln\rho A_2C_2C_3+8\rho C_2(2A_1A_2-4A_2c_0+
4A_2C_3-C_2^2+2)-16\ln\rho A_2^2C_3~~~~~~
$$
\begin{equation}
+8A_2(-2A_1A_2+2A_2c_0-2A_2C_3+C_2^2)=0. ~~~~~~~
~~ (C_1=0)~~~~~~~~~~~~~~~~~~~~
\end{equation}
In this way, expression (16) gives the logarithm solution
\begin{equation}
Y=c_0\rho^2\ln(\rho/\rho_0),
\end{equation}
when we choose that
$$
\lambda=\Delta p=C_1=C_2=A_2=0, ~ C_3=2c_0,
$$
\begin{equation}
C=-2c_0, ~ A_1=-2c_0\ln\rho_0+c_0.
\end{equation}
When $C_1=C_3=0$ and 
\begin{equation}
C=\frac{1}{2}A_1^2A_2-A_1A_2c_0-\frac{1}{4}A_1C_2^2+
A_2\Big(\frac{c_0^2}{2}+\frac{\lambda}{k}\Big)-c_0C_2^2,
\end{equation}
we obtained the solution
\begin{equation}
Y=\frac{A_1}{2}\rho^2-C_2\rho+A_2
\end{equation}
under the conditions of
\begin{equation}
A_1^2c_0-A_1c_0^2-\frac{2A_1\lambda}{k}-\frac{2\Delta p}{k}=0,
\end{equation}
\begin{equation}
C_2\Big(c_0^2+\frac{2\lambda}{k}-2A_1c_0\Big)=0,
\end{equation}
\begin{equation}
C_2(2A_1A_2-4A_2c_0-C_2^2+2)=0,
\end{equation}
\begin{equation}
A_2(2A_1A_2-2A_2c_0-C_2^2)=0.
\end{equation}
This solution contains the one presented in Ref. \cite{naito} 
due to the above conditions have the more general form. 
It degenerates to sphere, circular cylinders, tori 
and Delaunay's surface in the limiting cases. For instance, 
in the case of $A_2=C_2=0$, it reduces to the sphere 
solution with the condition $\Delta pr_0^2+2\lambda 
r_0-kc_0(2-c_0r_0)=0$ where $r_0=2/A_1$. 
And it gives the Delaunay's surface for $C_2=0$ 
and $A_2\not= 0$ under the condition  
$c_0\lambda+\Delta p=0$.

As the conclusions, on the basis of the discussion of 
the unique property for  the solution with finite 
terms about the general Helfrich shape equation, we change 
the general shape equation into a linear and a nonlinear 
differential equations. This equation system contains all 
known  rigorous solutions of the shape equation. 
We find also that the restricted conditions (22-25) 
of the polynomial solution (21). These conditions 
have the more general form comparing the known 
solutions \cite{naito}. 
These known solutions are all the particular 
solutions of a linear differential equation of the second 
orders. And the above discussions remind us that, except 
the solutions presented in this letter, the other  exact 
solutions which we guess it exists for the shape equation 
can not be denoted by the elementary functions. It is 
a challenged problem to find out them with the proper 
choice of the Ansatz. Another interesting 
problem is about the stability considerations of the new 
solution as in Ref. \cite{naito}. 
We hope the above discussions will 
be helpful for the study of biomembranes.

\end{document}